\documentclass[aps,physrev,amsmath,amssymb,twocolumn, superscriptaddress, 10pt, floatfix]{revtex4-2}
\usepackage{graphicx, xcolor}
\usepackage{microtype}
\usepackage{mathrsfs}
\usepackage{dsfont}
\usepackage[colorlinks,bookmarks=true,citecolor=blue,linkcolor=red,urlcolor=blue]{hyperref}
\hypersetup{
  pdftitle    = {Dissipative frequency converter: from Lindblad dynamics to non-Hermitian topology},
  pdfauthor   = {Florian Koch and Jan Carl Budich}
}
\usepackage{enumitem}
\usepackage{csquotes}

\usepackage[caption=false]{subfig}

\usepackage{physics}

\renewcommand{\H}{\mathcal{H}}
\renewcommand{\L}{\mathcal{L}}
\newcommand{\ui}{\mathrm{i}}
\newcommand{\symbf}[1]{\boldsymbol{#1}}
\DeclareMathOperator{\pur}{pur}

\begin{document}

\title{Dissipative frequency converter: from Lindblad dynamics to non-Hermitian topology}

\author{Florian Koch}
\email{florian.koch5@tu-dresden.de}
\affiliation{Institute of Theoretical Physics, Technische Universität Dresden and Würzburg-Dresden Cluster of Excellence ct.qmat, 01062 Dresden, Germany}
\author{Jan Carl Budich}
\email{jan.budich@tu-dresden.de}
\affiliation{Institute of Theoretical Physics, Technische Universität Dresden and Würzburg-Dresden Cluster of Excellence ct.qmat, 01062 Dresden, Germany}
\affiliation{Max Planck Institute for the Physics of Complex Systems, Nöthnitzer Str. 38, 01187 Dresden, Germany}
\date{\today}

\begin{abstract}
A topological frequency converter represents a dynamical counterpart of the integer quantum Hall effect, where a two-level system enacts a quantized time-averaged power transfer between two driving modes of incommensurate frequency. Here, we investigate as to what extent temporal coherence in the quantum dynamics of the two-level system is important for the topological quantization of the converter. To this end, we consider dissipative channels corresponding to spontaneous decay and dephasing in the instantaneous eigenbasis of the Hamiltonian as well as spontaneous decay in a fixed basis. The dissipation is modelled using both a full Lindblad and an effective non-Hermitian (NH) Hamiltonian description.
For all three dissipation channels we find a transition from the unperturbed dynamics to a quantum watchdog effect, which destroys any power transfer in the strong coupling limit.
This is striking because the watchdog effect leads to perfectly adiabatic dynamics in the instantaneous eigenbasis, at first glance similar to the unperturbed case.
Furthermore, it is found that dephasing immediately leads to an exponential decay of the power transfer in time due to loss of polarisation in the mixed quantum state.
Finally, we discuss the appearance in the effective NH trajectory description of non-adiabatic processes, which are suppressed in the full Lindblad dynamics.
\end{abstract}

\maketitle

\section{Introduction}
The influence of dissipation on physical phenomena is ubiquitous given that real physical systems are to some extent coupled to their environment \cite{Breuer2007,Rotter2015}.
Nevertheless, we tend to model idealized scenarios such as perfect isolation, so as to contain complexity and comprehensively explain observations with theory.
One of the most fascinating phenomena understood with mathematical precision for closed systems at zero temperature is the topological quantization of physical observables, such as the Hall conductance of a two-dimensional electron gas in a strong perpendicular magnetic field \cite{Klitzing1980,Laughlin1981,Thouless1982,Tsui1982,Halperin1982,Avron1983}.
The impressive accuracy to which theory and observation agree in this case has inspired a paradigm shift in the classification of phases of matter \cite{Zhang2005,Hasan2010,Klitzing2017,Sato2017}.
By now, the resulting notion of topological matter has been generalized from quantum materials to a broad range of physical settings \cite{Chiu2016,Xiao2010,Wen2017,Cayssol2021}.
Interestingly, going beyond closed systems and thermal equilibrium has not only revealed undesirable challenges to topological robustness, but has also led to the discovery of new topological phenomena without a direct zero-temperature counterpart \cite{Degen2017,Silberstein2020,Hoeckendorf2020,Budich2020,McDonald2020,Groenendijk2021,Koch2022}.

An intriguing example of  a genuinely non-equilibrium topological system is provided by a topological frequency converter \cite{Martin2017,Peng2018,Crowley2019,Nathan2019,Chen2020,Crowley2020,Koerber2020,Nathan2020,Long2021,Boyers2020,Malz2020,Koerber2022,Nathan2022,Schwennicke2022}.
There, a two-level system driven by two modes of incommensurate frequency enacts a quantized time-averaged power transfer between the two modes.
This is formally reminiscent of the aforementioned integer quantum Hall effect as both phenomena are governed by a topological invariant known as the first Chern number \cite{Martin2017,Xiao2010}.
However, the inherent long-time character of this phenomenon naturally raises the question as what extent quantum coherence is relevant to dynamical topological quantization.

\begin{figure}[htbp]
\includegraphics[scale=1.4]{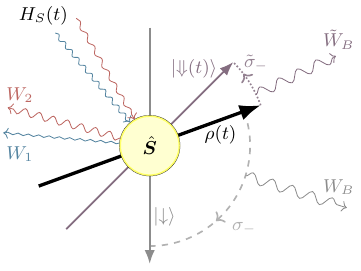}
\caption{
   Schematic illustration of the driven spin (yellow circle with $\hat{\symbf S}$) being affected by two different dissipation channels.
   The red and blue arrows on the left demonstrate the energy transfer $W_1, W_2$ provided by the topological frequency conversion due to the Hermitian Hamiltonian $H_S(t)$.
   The grey (purple-grey) arrows on the right illustrate the energy flow $W_B$ ($\tilde W_B$) from the system to a bath when the system state $\rho(t)$ spontaneously decays in a fixed $\sigma_z$-basis (in the instantaneous eigenbasis of $H_S(t)$) into the (instantaneous ground) state $\ket{\downarrow}$ ($\ket{\Downarrow}$), respectively.
}
\label{fig:illustration_model}
\end{figure}

Below, we study a dissipative extension of the topological frequency converter (DFC) in which the driven two-level system (or spin 1/2) is exposed to various sources of decoherence.
We investigate the temporal dynamics of the DFC both by solving a full quantum master equation \cite{Kossakowski1972,Lindblad1976} and within an effective non-Hermitian (NH) Hamiltonian approximation \cite{Ashida2020}.
In the absence of dissipation, topological quantization of the frequency transfer occurs in the slow driving limit, where the spin orientation adiabatically follows the Hamiltonian ground state, thus tracing out a state manifold with non-vanishing Chern number over time \cite{Martin2017}.
As a natural source of decoherence in this adiabatic limit, we consider both pure dephasing and relaxation of the spin in its instantaneous eigenbasis.
While any of the studied dissipative channels is found to immediately prevent a strict topological quantization, we observe a rich competition between coherent dynamics and decoherence.
For pure dephasing, the quantum state of the spin becomes increasingly mixed over time, thus leading to an exponential decay of the power transfer with a vanishing long-time average, independently of the dephasing strength.
By contrast, spin relaxation processes act as a repeated re-purification to the instantaneous ground state, thus enabling a finite long-time average of the power transfer.
Its magnitude only significantly deviates from the quantized value when the decay strength becomes comparable to the driving frequencies.
In the opposite limit of strong decay, both dephasing and relaxation exhibit a quantum watchdog effect \cite{Misra1977,Kraus1981} pinning the spin to its instantaneous ground state.
While this in some sense corresponds to perfectly adiabatic dynamics \cite{Born1928,Kato1950}, we find a vanishing power transfer in this scenario.
This observation highlights the fact that the response of the system is eventually carried by first dynamical corrections to the adiabatically evolved state \cite{Kubo1957,Kubo1957a,Weinberg2017,Koerber2022}.
While an effective NH Hamiltonian approach cannot capture pure dephasing, we are able to qualitatively and quantitatively confirm our findings for the spin relaxation channel there.
Finally, we consider dissipation in a constant $\sigma_z$-basis, for which the transition between weak and strong decoherence is accompanied by exceptional points (EPs) in the NH Hamiltonian \cite{Bergholtz2021}.
While the EPs themselves do not seem to have a striking effect on the power transfer, the concomitant complex energy landscape is found to strongly affect the onset of non-adiabatic processes.

\section{Model}
We consider a coherently driven spin that dissipatively interacts with its environment.
Within the Born-Markov approximation such open quantum systems are often described by Lindblad master equations \cite{Kossakowski1972,Lindblad1976,Breuer2007}
\begin{align}
    \dv{t}\rho = \L[\rho] = -\ui [\H, \rho] + \sum_\alpha \gamma_\alpha \left(L_\alpha \rho L_\alpha^\dagger - \frac{1}{2} \{\rho, L_\alpha^\dagger L_\alpha \} \right) \label{eq:Lindblad_Master_equation}
\end{align}
with the density matrix $\rho$, the system Hamiltonian $\H$, dissipation rates $\gamma_\alpha$ and the so-called jump operators $L_\alpha$.
In this work, we consider a time-dependent Hamiltonian $\H = H_S(t)$ in a slow (adiabatic) driving limit, slow as compared to the instantaneous level spacing \cite{Albash2012}.
In this regime, it is natural to assume the Lindblad dynamics with its instantaneous quantum jumps to be significantly faster than the external time-dependence of the Hamiltonian.
Along these lines, the Lindblad master equation (\ref{eq:Lindblad_Master_equation}) is assumed to be valid at all times, and the adiabatic time-dependence of the Hamiltonian may be reflected in slowly time-dependent jump operators $L_\alpha(t)$ that refer to the instantaneous eigenbasis of $H_S(t)$.

\subsection{Coherent topological frequency converter}
The Hamiltonian is given by the topological frequency converter introduced in Ref.~\cite{Martin2017}, modeling a spin coupled to a time-dependent quasi-periodic magnetic field $\symbf{B}(\vec\varphi_t)$, i.e.
\begin{align}
H_S(t) = g^* \mu_B \symbf{B}(\vec\varphi_t) \cdot \hat{\symbf S}
\label{eq:tfc}
\end{align}
where $\hat{\symbf S} = \frac{1}{2} \symbf{\sigma}$ represents the spin-$1/2$, $\mu_B$ is Bohr's magneton and $g^*$ is the effective $g$-factor of the spin.
The time-dependent magnetic field $\symbf B(\vec \varphi_t) = B_c \symbf d(\vec\varphi_t)$ reads as
\begin{align}
   \symbf{d}(\vec\varphi_t) =
        \begin{pmatrix}
            \sin{\varphi_{1t}} \\
            \sin{\varphi_{2t}} \\
            m - \cos{\varphi_{1t}} - \cos{\varphi_{2t}}
        \end{pmatrix}
\end{align}
with a static magnetic field $B_s = B_c m$ in the $z$-direction and two circularly polarized drives with amplitudes $B_c$ and time-dependent phases $\vec\varphi_t = (\varphi_{1t}, \varphi_{2t}) = \vec\omega t + \vec\phi$.
The offset phases and incommensurate frequencies are parameterized by $\vec\phi = (\phi_1, \phi_2)$ and $\vec\omega = (\omega_1, \omega_2)$.
This model may be regarded as a temporal version of the half BHZ model, introduced by Bernevig, Hughes and Zhang, which is usually considered as a standard Chern insulator in two spatial dimensions \cite{Bernevig2006}.
Here, the parameters $\vec k$ are replaced with the time-dependent phases $\vec\varphi_t$.
Similar to Ref.~\cite{Martin2017} we combine the prefactors to $\eta = \frac{1}{2} g^* \mu_B B_c$ such that the Hamiltonian can be written as $H_S(t) = \eta (\symbf{d} \cdot \symbf\sigma)$ where $\eta$ describes the overall energy scale.

In Ref.~\cite{Martin2017} it was shown that this model in the slow-driving limit (described by $H_S(t) = \symbf h_1(\varphi_{1t}) \cdot \symbf\sigma + \symbf h_2(\varphi_{2t}) \cdot \symbf\sigma$ with $\symbf h_1(\varphi_{1t}) = \eta(\sin(\varphi_{1t}), 0, \frac{m}{2} - \cos(\varphi_{1t}))^T$ and $\symbf h_2(\varphi_{2t}) = \eta(0, \sin(\varphi_{2t}), \frac{m}{2} - \cos(\varphi_{2t}))^T$) exhibits an average energy transfer between the two driving modes written as
\begin{align}
   \frac{1}{2}\pdv{(E_1 - E_2)}{t} = \abs{\Omega_{\vec q}} \omega_1 \omega_2. \label{Eq:energy_transfer_Berry_curv}
\end{align}
Here $E_i = \symbf h_i(\varphi_{it}) \cdot \expval{\symbf \sigma}$ are the individual beam energies, $\vec q(t) = \vec\omega t$ and $\Omega_{\vec q} = \hat z i \frac{1}{2}\tr\left(P_{\vec q} \left[\pdv{P_{\vec q}}{q_1}, \pdv{P_{\vec q}}{q_2}\right]\right)$ is the Berry curvature with $P_{\vec q}$ a projector onto the instantaneous ground state parameterized by $\vec q(t)$ \cite{Sakurai2020}.
Using Floquet theory for multiple driving frequencies \cite{Floquet1883,Ho1983}, this model can be mapped into a two-dimensional tight-binding model with an intrinsic electric field $\vec E = \vec \omega$, where the time evolution of the spin can be understood as a hopping process on the so-called Floquet lattice \cite{Nakanishi1993}. Note that these two extra dimensions of the spatially zero-dimensional two-level system only emerge by mathematical analogy.

Semiclassical equations of motion \cite{Sundaram1999} connect the energy transfer to the Berry curvature (see Eq.~(\ref{Eq:energy_transfer_Berry_curv})) in a way that is analogous to the Hall response in a two-dimensional electron gas \cite{Xiao2010}.
Furthermore, since the spin homogeneously probes the entire Floquet-Brillouin zone over time, the time-averaged response is given by $\overline{\Omega_{\vec q}} = \int \frac{\dd[2]{q}}{(2\pi)^2} \Omega_{\vec q} = \frac{\mathcal C}{2\pi}$ where $\mathcal C$ is the first Chern number \cite{Chern1946,Sakurai2020}.
Thus, the average energy transfer is indeed topologically quantized and proportional to the Chern number of the synthetic 2D Floquet lattice space by analogy to the integer quantum Hall effect \cite{Martin2017}.

\subsection{Dissipative generalization and energy transfer}
In this work, we extend the topological frequency converter model (\ref{eq:tfc}) by introducing dissipative channels so as to arrive at a Lindblad master equation of the form (\ref{eq:Lindblad_Master_equation}).
In particular, we study spontaneous decay and dephasing in the eigenbasis of the Hamiltonian as well as spontaneous decay in a fixed $\sigma_z$ basis.
Since $H_S(t)$ is Hermitian, it is diagonalizable with
\begin{align}
H_S(t) = U(t)\cdot D(t) \cdot U^\dagger(t)
\label{eq:hsdiag}
\end{align}
with a unitary matrix $U(t)$ and a real-valued diagonal matrix $D(t) = \alpha(t) \mathds{1} + \beta(t) \sigma_z$.
Thus, the Pauli matrices in the basis of the instantaneous Hamiltonian are given by
\begin{align}
\tilde\sigma_i(t) = U(t)\cdot\sigma_i \cdot U^\dagger(t).
\label{eq:sigtilde}
\end{align}
We denote the two instantaneous eigenstates of the Hamiltonian by $\ket{{\Uparrow}}$ and $\ket{{\Downarrow}}$ where $\Uparrow(\Downarrow)$ indicates the upper (lower) band of the Hamiltonian.
The double arrows are used to visually distinguish the (time-dependent) instantaneous eigenstates from the usual $\sigma_z$ eigenstates $\ket{\uparrow}$ and $\ket{\downarrow}$ in a fixed basis.
The three considered dissipation channels are modeled by the jump operators $L_1 = \tilde\sigma_-(t)$ with $\tilde\sigma_-(t)\ket{{\Downarrow}} = 0$ and $\tilde\sigma_-(t)\ket{{\Uparrow}} = \ket{{\Downarrow}}$, $L_2 = \tilde\sigma_z(t)$ with $\tilde\sigma_z(t) \ket{{\Downarrow}} = -\ket{{\Downarrow}}$ and $\tilde\sigma_z(t) \ket{{\Uparrow}} = \ket{{\Uparrow}}$, and $L_3 = \sigma_-$ with $\sigma_- \ket{\downarrow} = 0$ and $\sigma_- \ket{\uparrow} = \ket{\downarrow}$.

While the dissipative channels influence the time evolution, they generally also draw energy from the system \cite{Breuer2007}.
For this reason, we introduce a bath current $W_B$ using the common open quantum system approach, i.e.~we separate the energy flows of the Lindblad master equation into internal (taking place in the system) and external (system-bath coupling)
\begin{align}
   \dv{\expval{E}}{t}
   = \dv{t} \tr[H_S \rho]
   = \underbrace{\tr[\dot H_S \rho]}_{\text{internal}} + \underbrace{\tr[H_S \dot\rho].}_{\text{external}}
\end{align}
The internal current is divided into two energy currents corresponding to the given sources.
Consider $H_S(t) = \symbf h_1(\varphi_{1t}) \cdot \symbf\sigma + \symbf h_2(\varphi_{2t}) \cdot \symbf\sigma$.
Then (omitting the arguments for readability and brevity)
\begin{align}
   \tr[\dot H_S \rho] &= \tr\left[\left(\pdv{\symbf h_1}{t} \cdot \symbf\sigma + \pdv{\symbf h_2}{t} \cdot \symbf\sigma\right) \rho \right] \\
   &= \pdv{\symbf h_1}{t} \cdot \expval{\symbf\sigma} + \pdv{\symbf h_2}{t} \cdot \expval{\symbf\sigma} \\
   &\equiv \dv{W_1}{t} + \dv{W_2}{t}
\end{align}
The two terms $\dv{W_1}{t}$ and $\dv{W_2}{t}$ are interpreted as the energy change due to one of the energy sources \cite{Martin2017}.

The external current $\dv{W_B}{t} = \tr[H_S \dot\rho]$ induced by the bath depends on the time evolution of the system $\dot\rho$.
Here we have three different cases:
\begin{enumerate}[label=(\roman*)]
  \item Hermitian time evolution ($\gamma_\alpha = 0$) governed by the von Neumann equation $\dot\rho = -\ui [H_S, \rho]$ \cite{Sakurai2020} leads to $\tr[H_S \dot\rho] = 0$, i.e.~no current to the bath.
  So the Hermitian case is exactly the case treated in Ref.~\cite{Martin2017}.
  \item Liouvillian time evolution governed by the Lindblad master equation $\dot\rho = \mathcal L[\rho]$ (see Eq.~(\ref{eq:Lindblad_Master_equation})) leads to non-vanishing current $\dv{W_\text{B}^\text{L}}{t} = \sum_\alpha \gamma_\alpha \left(\tr[L_\alpha^\dagger H_S L_\alpha \rho] - \frac{1}{2} \tr[\{H_S, L_\alpha^\dagger L_\alpha\} \rho] \right)$ and
  \item Non-Hermitian time evolution governed by the non-linear differential equation (\ref{eq:non-Hermitian_equation}) leads to a non-vanishing current of the form $\dv{W_\text{B}^\text{NH}}{t} = \sum_\alpha \gamma_\alpha \left(\tr[L_\alpha^\dagger L_\alpha \rho] \tr[H_S \rho] - \frac{1}{2} \tr[\{H_S, L_\alpha^\dagger L_\alpha\} \rho]\right)$.
\end{enumerate}
The time integral $W_i(t) = \int_0^t \dv{W_i}{t}(t') \dd{t'}$ is the total work performed by a drive after time $t$.
With these definitions, we can derive a Kirchhoff-type law for the currents from the system dynamics.
We find that the averaged sum of the currents over a longer time $t$ goes to zero
\begin{align}
   \lim_{t\to\infty} \frac{W_1(t) + W_2(t) + W_B(t)}{t} = 0
   \label{eq:Kirchhoff}
\end{align}
This law only holds in the infinite time limit because the spin acts as a memory for a small amount of energy, which naturally leads to small temporary deviations from $W_1(t) + W_2(t) + W_B(t) = 0$.

We also define the total effective energy transferred by
\begin{align}
   \Delta E = |W_1(t) - W_2(t)| - |W_B(t)|. \label{eq:effective_energy_transfer}
\end{align}
Note that the term \emph{energy transfer} only makes sense if the signs of the two internal energy flows are opposite because otherwise, the two energy beams just transfer energy to the bath. In the coherent case, this quantity satisfies $\pdv{t} \Delta E = 2 |\Omega_{\vec q}| \omega_1 \omega_2$ with $\overline{\Omega_{\vec q}} = \int \frac{\dd[2]{q}}{(2\pi)^2} \Omega_{\vec q} = \frac{\mathcal C}{2\pi}$, such that the time-averaged (denoted by the overline) effective energy transfer $\overline{\partial_t (\Delta E)}$ is topologically quantized \cite{Martin2017,Sakurai2020}.

\subsection{Effective non-Hermitian Hamiltonian}
Finally, we introduce the effective non-Hermitian approximation to the Lindblad master equation (Eq.~(\ref{eq:Lindblad_Master_equation})), which naturally emerges from the quantum trajectory picture \cite{Griffiths1993,Wiseman1996,Daley2014,Ashida2020}.
There, the Lindblad master equation is understood as an average of stochastically evolved wavefunctions that include so-called quantum jumps.
Defining the effective non-Hermitian Hamiltonian as
\begin{align}
   H_\text{NH} = H_S - \ui \Lambda \label{eq:NH_Hamiltonian}
\end{align}
with $\Lambda = \frac{1}{2} \sum_\alpha \gamma_\alpha L_\alpha^\dagger L_\alpha$ allows us to rewrite the Lindblad master equation as
\begin{align}
   \dv{t}\rho = \L[\rho] = -\ui\left( H_\text{NH} \rho - \rho  H^\dagger_\text{NH}\right) + \sum_\alpha \gamma_\alpha L_\alpha \rho L_\alpha^\dagger.
\end{align}
For a (pure) state $\ket{\psi}$ one naively identifies $P_\alpha(\Delta t) = \langle\psi | \gamma_\alpha L^\dagger_\alpha L_\alpha | \psi\rangle \Delta t$ as the probability of a jump with $L_\alpha$ occurring in the short time-interval $\Delta t$.
This immediately allows the stochastic evolution of the system state by propagating with the effective non-Hermitian Hamiltonian ($\ket{\psi(t + \Delta t)} = \frac{1}{\mathcal{N}} \exp(-\ui H_\text{NH} \Delta t) \ket{\psi(t)}$) until a quantum jump occurs $\ket{\psi(t + \Delta t)} = \frac{1}{\mathcal{N}} L_\alpha\ket{\psi(t)}$ \cite{Griffiths1993,Wiseman1996,Daley2014,Ashida2020}.
The probability of such a transition appearing depends on the non-Hermitian time evolution.
Especially, an increase in the coupling strength usually comes with an increase of those sudden transitions.
In this sense, the evolution according to the effective non-Hermitian Hamiltonian describes a quantum trajectory conditional on no occurring jumps.
The comparison between this non-Hermitian and the full Lindbladian dynamics may thus provide insights into the importance of the actual quantum jump processes.

To keep the states during the non-Hermitian time evolution normalized, one can either renormalize the state after each incremental time step (see normalization factor $\mathcal{N}$), or, equivalently, use the nonlinear equation of motion \cite{Sergi2019}
\begin{align}
   \dv{t}\rho = -\ui [H_S, \rho] - \{\Lambda, \rho\} + 2 \rho \tr[\Lambda\rho]. \label{eq:non-Hermitian_equation}
\end{align}

\section{Methods}
In the following section, we will give a brief summary of the numerical approaches used to solve the Lindblad master equation and state the initial conditions of the simulation.
Afterwards, we introduce the Fidelity as a measure of how close two states are.
This will allow us to quantify  the drastic changes that happen in the system due to the increase of dissipation.

\subsection{Numerical approach to dissipative dynamics}
\label{ssec:numerical_approach}
The Lindblad master equation (\ref{eq:Lindblad_Master_equation}) is a linear differential equation in the density matrix $\rho$, even though the jump operators $L_\alpha$ act on the density matrix $\rho$ from both sides.
By vectorizing the density matrix $\rho \to \vec r$, we may transform the Lindblad master equation into the usual form $\dv{t} \vec r = L \cdot \vec r$ of a linear equation.
Here, the so-called Lindblad superoperator $\mathcal L$ is transformed into a square matrix $L$ which acts on the density matrix vector $\vec r$ from the left only \cite{Machnes2014, Horn1994}.
This well-known procedure can be used to solve open systems with a small Hilbert space.
In the system presented, the 2x2 density matrix $\rho$ is transformed into a vector $\vec r$ of length 4 and the Lindblad superoperator $\mathcal{L}$ into a 4x4 matrix $L$.

Since the investigated system is quasiperiodically driven, the matrix $L$ is time-dependent and therefore cannot be solved by simple exponentiation.
A standard method to solve this kind of equations is given by commutator-free exponential time propagators, also known as Magnus expansion.
Here we used an optimized fourth-order Magnus expansion (see Eq. (21) in Ref.~\onlinecite{Alvermann2012}).
The Magnus expansion in combination with a step-wise renormalisation of the wave function was also used for the numerical evaluation of the effective NH Hamiltonian dynamics.

To allow for direct comparison with the  coherent  topological frequency converter, we use the same parameters for the Hermitian Hamiltonian as in Ref.~\onlinecite{Martin2017}, i.e.~the incommensurate frequencies are given by $\omega_1 = 0.1$ and $\frac{\omega_2}{\omega_1} = \frac{1}{2}(\sqrt{5} + 1) \approx 1.618$, the golden ratio, and the offset phases are chosen $\phi_1 = \frac{\pi}{10}$ and $\phi_2 = 0$, and the overall energy scale $\eta = 2$.
The time evolution is performed with an incremental time step of $\Delta t = 0.01$ up to times $t_\text{max} = 10^4$.
In sections \ref{sec:inst_eigenbasis} and \ref{sec:fixed_basis} we present the results for $m = 1$, where the system is in the topological region characterized by the Chern number $\mathcal{C} = +1$.
For other values of $m$ within the topologically non-trivial region $0 < |m| < 2$, we obtain qualitatively similar results (cf.~Appendix~\ref{app:results_decay_topology}).
In the topologically trivial region, no energy transfer occurs.
To obtain fast convergence to the long-time dynamics, we used the instantaneous ground state as initial state $\rho(0) = \ket{\Downarrow}\bra{\Downarrow}$.

\subsection{Fidelity as a diagnostic tool}
A well-known quantity often used in quantum information theory to measure how \enquote{close} two states are is the Fidelity \cite{Jozsa1994}.
For the states $\rho$ and $\sigma$ it is defined as
\begin{align}
   F(\rho, \sigma)
   &= \left(\tr\sqrt{\sqrt{\rho} \sigma \sqrt{\rho}}\right)^2.
\intertext{For qubits this simplifies to}
     F(\rho, \sigma) &= \tr[\rho\sigma] + 2 \sqrt{\det[\rho] \det[\sigma]}. \label{eq:Fidelity}
\intertext{Further, if one of the states is pure $\sigma = \ket{\psi}\bra{\psi}$ then $\det[\sigma] = 0$ and the Fidelity reduces to}
     F(\rho, \sigma) &= \bra{\psi} \rho \ket{\psi}.
\end{align}
The Fidelity can take values between $0$ and $1$ with $F(\rho, \sigma) = 1$ if $\rho = \sigma$ \cite{Jozsa1994}.

Further, we define the \emph{projected Fidelity} $F_\text{p}(\rho, \sigma)$ as the fidelity of the (mixed) states $\rho$ and $\sigma$ projected onto their respective pure states $\rho \to \frac{1}{2}\left(\mathds{1} + \tilde{\symbf d}\cdot \symbf\sigma\right)$ with $\tilde{\symbf d} = \frac{\symbf{d}}{\abs{\symbf d}}$.
In the picture of the Bloch sphere, the respective pure states are the projections of the states onto the surface of the Bloch sphere.
This quantity allows us to measure the \enquote{closeness} of any mixed state $\rho$ with a pure state $\ket{\psi}$ independent of the purity of the mixed state, as the fidelity can be written as $F(\rho, \ket{\psi}) = \abs{\symbf d} F_\text{p}(\rho, \ket{\psi}) + \frac{1}{2} (1 - \abs{\symbf d})$.
This construction will allow for deeper insights into the effect of the Lindblad jump operators on the energy transfer.

\section{Decoherence in instantaneous eigenbasis}
\label{sec:inst_eigenbasis}
In this section, we discuss the DFC with decoherence in the instantaneous eigenbasis of the Hamiltonian (cf.~Eq.~(\ref{eq:sigtilde})), both in the full Liouvillian description and the effective NH approximation.
We model spontaneous decay and dephasing with the Lindblad jump operators $L_-(t) = \tilde\sigma_-(t)$ and $L_z(t) = \tilde\sigma_z(t)$, respectively.

\subsection{Full Liouvillian description}
Before we dive into the discussion of our results, we give a brief discussion of the two types of decoherence.
In the quantum trajectory picture \cite{Griffiths1993,Wiseman1996,Daley2014,Ashida2020}, the action of the jump operators lead to sudden transitions $\rho(t+\dd{t}) = L_\alpha(t) \rho(t) L_\alpha^\dagger(t)$ up to normalization, and the probability of such a transition is governed by the NH time evolution encompassing the dissipation rates $\gamma_\alpha$.
Specifically, spontaneous decay leads to $\rho(t+\dd{t}) = \ket{\Downarrow}\bra{\Downarrow}$, i.e.~a projection to the instantaneous ground state. By contrast, dephasing results in $\rho(t+\dd{t}) = c_1 \ket{\Downarrow}\bra{\Downarrow} + c_2 \ket{\Uparrow}\bra{\Uparrow}$, thus destroying all the off-diagonal elements in the instantaneous eigenbasis while leaving the diagonal entries untouched.
With increasing dissipation rates $\gamma_\alpha$ these jumps are happening more often, leading to a transition from a regime where the coherent motion of the spin is not altered significantly (small coupling limit) to a regime where we find signatures of a quantum watchdog effect (large coupling limit), i.e.~the state becomes pinned to the target state $L_\alpha(t) \rho(t) L_\alpha^\dagger(t)$.
Since the target state is the ground state of the spin, we may consider this pinning to the instantaneous eigenstates as supporting an ideal evolution according to the adiabatic theorem \cite{Born1928}. Remarkably, as we demonstrate below, the power transfer between the two driving modes completely vanishes in this strong coupling regime, despite the ideal adiabatic evolution tracing out topologically non-trivial state manifold.

\begin{figure}[htbp]
\includegraphics{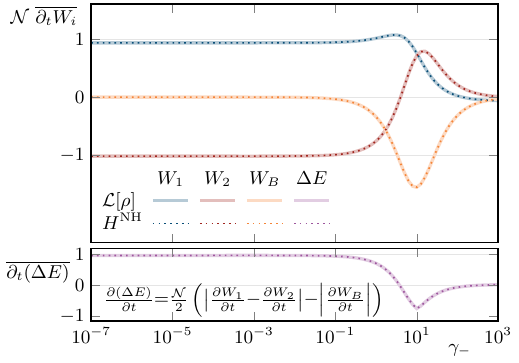}
\caption{
   Energy currents as a function of dissipation rate $\gamma_-$ for spontaneous decay in the instantaneous eigenbasis of $H_S(t)$ (see Eq.~(\ref{eq:tfc})).
   Top: Temporal average of the energy and bath currents $\mathcal{N}\pdv{W_i}{t}$ with $\mathcal{N} = \frac{2\pi}{\omega_1 \omega_2}$.
   Bottom: Temporal average of the effective energy transfer $\pdv{(\Delta E)}{t} = \frac{\mathcal{N}}{2} |\pdv{W_1}{t} - \pdv{W_2}{t}| - |\pdv{W_B}{t}|$.
   Note the accurate agreement between full Liouvillian (solid lines) and effective non-Hermitian dynamics (dotted lines) in all shown plots.
   For further parameter choices see Section~\ref{ssec:numerical_approach}.
}
\label{fig:currents_eigenbasis_both}
\end{figure}

\begin{figure}[htbp]
\includegraphics{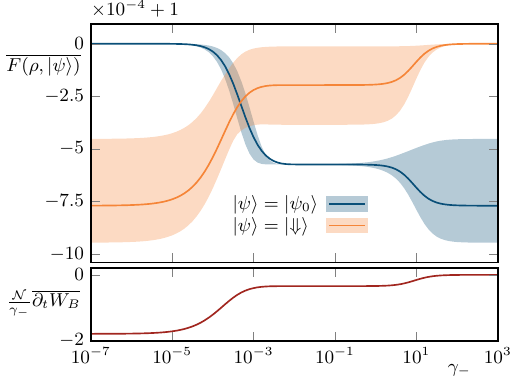}
\caption{
   Fidelities and the $\gamma_-$-normalized bath current as functions of the dissipation rate $\gamma_-$ for spontaneous decay in the instantaneous eigenbasis of $H_S(t)$ (see Eq.~(\ref{eq:tfc})).
   Top: Temporal averaged fidelities (and $0.25$- and $0.75$-quantils) of the state $\rho(t)$ with unperturbed state $\ket{\psi_0}$ and with instantaneous eigenstate of the Hamiltonian $\ket{\Downarrow}$.
   Bottom: Temporal average of $\gamma_-$-normalized bath current $\frac{\mathcal{N}}{\gamma_-} \pdv{W_B}{t}$.
   Note the simulationeous transitions from $\rho(t) = \ket{\psi_0}\bra{\psi_0} \to \ket{\Downarrow}\bra{\Downarrow}$ and $\frac{\mathcal{N}}{\gamma_-} \overline{\partial_t W_B} \to 0$.
}
\label{fig:results_eigenbasis_Lindblad}
\end{figure}

\begin{figure}[htbp]
\includegraphics{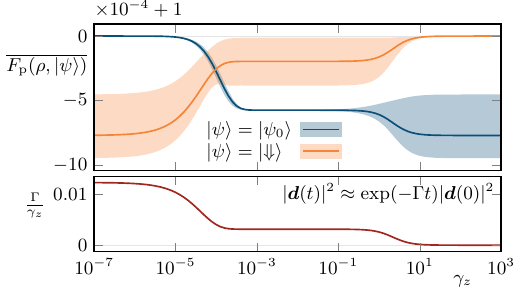}
\caption{
   Projected fidelities and the $\gamma_z$-normalized purity decay exponent as functions of dissipation rate $\gamma_z$ for dephasing in the instantaneous eigenbasis of $H_S(t)$ (see Eq.~(\ref{eq:tfc})).
   Top: Temporal average of the projected fidelities (and $0.25$- and $0.75$-quantils) of the state $\rho(t)$ with unperturbed state $\ket{\psi_0}$ and with instantaneous eigenstate of the Hamiltonian $\ket{\Downarrow}$.
   Bottom: $\gamma_-$-normalized purity decay exponent $\frac{\Gamma}{\gamma_z}$ with $\abs{\symbf d(t)}^2 \approx \exp(-\Gamma t) \abs{\symbf d(0)}^2$ (see Eq.~(\ref{eq:Gamma})).
}
\label{fig:results_Dephasing}
\end{figure}

In order to study the transition from weak to strong dissipation, we calculate the currents $W_1(t), W_2(t)$ and $W_B(t)$, as well as the fidelity of the system state $\rho(t)$ with both the instantaneous groundstate of the Hamiltonian $\ket{\Downarrow}$ and the state $\ket{\psi_0}$ of an unperturbed (i.e.~coherently evolving) system.
For spontaneous decay, the bath current takes the form
\begin{align}
  \dv{t} W_B^\text{L}
  = -2\gamma (1 - F(\rho, \ket{\Downarrow})) \beta(t)
\end{align}
with $2\beta(t) = \tr[H_S \tilde\sigma_z] = \expval{H_S}{\Uparrow} - \expval{H_S}{\Downarrow}$ the energy splitting between the instantaneous eigenstates.
For the system state being exactly the instantaneous eigenstate $\rho(t) = \ket{\Downarrow}\bra{\Downarrow}$ (perfectly adiabatic) this evaluates to $W_B^\text{L} = 0$.
The dephasing channel does not consume energy from the system, since $\dv{t} W_\text{B}^\text{L} = 0$.
The results for spontaneous decay are illustrated in Figs.~\ref{fig:currents_eigenbasis_both} and \ref{fig:results_eigenbasis_Lindblad} and the results for dephasing in Fig.~\ref{fig:results_Dephasing}.

In Fig.~\ref{fig:currents_eigenbasis_both} we see the slopes of the three currents $W_1(t), W_2(t)$ and $W_B^\text{L}(t)$ (solid lines) as the decay rate $\gamma_-$ increases.
The slopes appear to be near constant at the topologically quantized value up to $\gamma_- \approx 10^{-1}$, where the bath current starts to increase significantly, until it vanishes again at large $\gamma_-$.
Due to the approximate Kirchhoff-type law (see Eq.~(\ref{eq:Kirchhoff})), the sum of the energy currents $W_1(t) + W_2(t)$ behaves in a similar way but with opposing sign.
Interestingly, the individual currents behave differently, as $W_1(t)$ vanishes almost solely and $W_2(t)$ peaks and even changes its sign before it also vanishes.
With the change of sign of $W_2(t)$, there is no more energy transfer between the driving modes, and all energy is transferred to the bath.

Fig.~\ref{fig:results_eigenbasis_Lindblad} will help interpreting the bare results of Fig.~\ref{fig:currents_eigenbasis_both} and connect the insights with the aforementioned transition from weak to strong coupling.
In particular, in Fig.~\ref{fig:results_eigenbasis_Lindblad}, the slope of $\frac{\mathcal N}{\gamma_-} W_B^\text{L}(t)$ and the mean fidelities with the $(0.25, 0.75)$ quantiles of the state $\rho(t)$ with the unperturbed state $\ket{\psi_0}$ as well as the instantaneous ground state $\ket{\Downarrow}$ are plotted over $\gamma_-$.
At first glance, one can recognize a similar behavior of all three curves, albeit the Fidelity of $\rho(t)$ with the instantaneous ground state being upside down.
In more detail, we can see that the slope of $\frac{\mathcal N}{\gamma_-} W_B^\text{L}(t)$ decreases in three steps from around $1.75$ to $0.35$ to $0$.
Including the two fidelities into our analysis, we can see that these steps are accompanied by a transition from $\rho(t) \approx \ket{\psi_0} \bra{\psi_0}$ to an intermediate state to $\rho(t) \approx \ket{\Downarrow}\bra{\Downarrow}$.
This behavior is exactly the transition described above from a regime where the adiabatic dynamics of the spin are not altered by the coupling to the bath to a regime where we find signatures of a quantum watchdog effect.

Returning to Fig.~\ref{fig:currents_eigenbasis_both}, we see that the behavior of the currents up to their peaks is explained by the proportionality to $\gamma_-$ with an exponential increase shown in logarithmic scale.
Further, the fact that all three currents vanish at even stronger dissipation can be explained with the watchdog effect.
For this, we calculated $W_1(t)$ and $W_2(t)$ specifically for the hypothetical ideally adiabatic state $\rho(t) = \ket{\Downarrow}\bra{\Downarrow}$, where we find that both currents vanish individually.
In other words, although the spin being pinned to the instantaneous ground state of the Hamiltonian at all times, not only the bath current vanishes but also the power transfer which would be naively expected to exhibit perfect topological quantization for perfect adiabatic dynamics. This observation highlights the fact that the first non-adiabatic deviations are a necessity for the appearance of the topologically quantized energy transfer, even though the corresponding topological invariant (Chern number) is defined by the adiabatic state manifold.

Next, we describe our findings for the DFC with pure dephasing, where we find both similarities and key differences to the spontaneous decay.
In particular, we will see that there are also signatures of a watchdog effect which destroys the energy transfer.
However, the main difference to the spontaneous decay is the dynamics of the purity of the quantum state.
As the target state of the spontaneous decay is the pure groundstate of the Hamiltonian, there the system state $\rho(t)$ remains a pure state in the long time limit (see Appendix~\ref{app:purity_decay}).
For pure dephasing, by contrast, each quantum jump reduces the purity of the state (see Appendix~\ref{app:purity_dephasing}) which leads to the convergence of the system state to the fully mixed state $\rho(t) \to \frac{1}{2}\mathds{1}$ in the long time limit. Thus, independent of the coupling strength $\gamma_z$, pure dephasing leads to a vanishing long-time average of the power transfer.

In particular, we find that the length of the state Bloch vector $\symbf{d}(t)$ is monotonically decreasing
\begin{align}
   \dv{t} \abs{\symbf d(t)}^2
   = 4\gamma_z \abs{\symbf d(t)}^2 \left[\cos\measuredangle(\symbf d(t), \symbf\beta(t)) - 1\right] \leq 0 \label{eq:ddt_pur}
\end{align}
with $\symbf\beta(t)$ the \emph{Bloch vector} of the Hamiltonian, i.e.~$H(t) = \alpha(t) + \symbf\beta(t) \cdot \symbf\sigma$.
From this differential equation we can see that
\begin{align}
   \frac{\abs{\symbf d(t)}^2}{\abs{\symbf d(0)}^2}
   = \exp(4\gamma_z \int_0^t \left[\cos\measuredangle(\symbf d(t'), \symbf \beta(t')) - 1\right] \dd{t'})
\end{align}
and thus $\exp(-8\gamma_z t) \leq \frac{\abs{\symbf d(t)}^2}{\abs{\symbf d(0)}^2} \leq 1$.
Approximating the angle of the vectors by an mean angle $\measuredangle(\symbf d(t'), \symbf \beta(t')) \approx \overline{\measuredangle(\symbf d, \symbf \beta)}$ gives
\begin{align}
   \abs{\symbf d(t)}^2 \approx \exp(-\Gamma t) \abs{\symbf d(0)}^2 \label{eq:Gamma}
\end{align}
with $\Gamma = 4\gamma_z \left[1 - \cos\overline{\measuredangle(\symbf d, \symbf \beta)}\right]$.
Numerically, we find that this approximation describes the behavior of $\abs{\symbf d(t)}$ very well.

As the ground state and the exited state have opposing Chern numbers $\mathcal C = \pm 1$ and $\rho(t) \to \frac{1}{2}\mathds{1}$ may be understood as a classical mixture of the two energy bands, the energy currents $W_1(t)$ and $W_2(t)$ should behave as $W_i(t) \approx \frac{2\alpha_0}{\Gamma} \left(1 - \exp(-\frac{\Gamma}{2}t)\right)$ where $\alpha_0$ is the slope of the unperturbed system.
Interestingly, this formula does not give the full picture of the dephasing as it does not catch the aforementioned watchdog effect.

In Fig.~\ref{fig:results_Dephasing}, we illustrate the fit of the exponent $\frac{\Gamma}{\gamma_z}$ and, similar to the case of spontaneous decay, the (projected) Fidelities with its $(0.25, 0.75)$-quantiles of the system state $\rho(t)$ with both, the unperturbed state $\ket{\psi_0}$ and the ground state $\ket{\Downarrow}$ over the dephasing strength $\gamma_z$.
One can see that the projected fidelities behave similarly to the fidelities of the spontaneous decay.
Thus, irrespective of the purity of $\rho(t)$, we can see a transition from the unperturbed state to an intermediate state to the ground state of the system.

Furthermore, the purity decay exponent $\frac{\Gamma}{\gamma_z}$ behaves similar to the slope of the bath current $\frac{\mathcal N}{\gamma_-} W_B^\text{L}$ as it goes stepwise from around $0.012$ to $0.003$ to $0$.
Interestingly, $\Gamma$ being close to zero indicates that the system state $\rho(t)$ stays pure over time, which, at first glance, gives hope for a non-vanishing energy transfer.
However, as this regime corresponds to the aformentioned watchdog effect $\rho(t) \approx \ket{\Downarrow}\bra{\Downarrow}$, no power transfer is possible there.

From above analysis we can conclude that both spontaneous decay and dephasing exhibit a quantum watchdog effect in the strong coupling limit where the system state $\rho(t)$ gets pinned to the instantaneous ground state $\ket{\Downarrow}$ of the Hamiltonian.
We have shown that this state does not support any power transfer indicating that the topological response is carried by first dynamical corrections to the adiabatically evolved state.
In addition, we find that spontaneous decay, in contrast to dephasing, allows for the long-time average of the power transfer to be finite and arbitrarily close to the topologically quantized coherent value at weak dissipation.
With dephasing, this is prohibited due to the continuous reduction of the purity of $\rho(t)$.
Comparing the dephasing case with the discussion in Ref.~\onlinecite{Hu2016}, we find a major difference between systems with two spatial dimensions and our present Floquet-lattice dimensions.
Specifically, for a 2D Chern insulator dephasing not only does not destroy the Hall response, but may actually stabilize it \cite{Hu2016}.
This effect is due to the locality of the dephasing in the 2D Chern insulator. There, dephasing only significantly affects states with $k_x, k_y$ near the Dirac point.
In our present two-frequency driving case, however, the reduction of the purity (shrinking of the Bloch sphere) affects the state at all times.
Furthermore, purity cannot be restored by the Hamiltonian evolution, and thus inevitably $\rho \to \frac{1}{2} \mathds{1}$.

When both dephasing and spontaneous decay in the instantaneous eigenbasis are present, we find that the long-time averaged power transfer between the two driving modes survives the onset of dephasing since the dephasing-induced loss of purity (or polarization) of the quantum two-level system is reversed by each spontaneous decay event. For quantitive numerical data regarding the interplay of the two dissipation channels, we refer the reader to  Appendix \ref{app:results_gamma_md_d}.

\subsection{Effective non-Hermitian description}
The effective non-Hermitian Hamiltonian is given by Eq.~(\ref{eq:NH_Hamiltonian}), where the anti-Hermitian part for spontaneous decay reads $\Lambda = \frac{1}{4} \gamma_- \left(\mathds{1} + \tilde\sigma_z(t)\right)$ and for dephasing $\Lambda = \frac{1}{2} \gamma_z \mathds{1}$.
Since a uniform damping term ($\Lambda\propto \mathds{1}$) does not alter the (conditional on no jump) dynamics up to normalization, it can be ignored.
Hence, the effective NH description of the dephasing channel does not change the results of the unperturbed topological frequency converter, and thus cannot even capture the basic physical phenomenology of the full dephasing dynamics studied above.
In particular, the quantized power transfer is preserved here indicating that both the loss of purity and the watchdog effect solely arise in the full Liouvillian description that includes quantum jumps.

For the case of spontaneous decay, after subtracting the aforementioned overall imaginary energy shift the effective NH Hamiltonian reads as (cf.~Eq.~(\ref{eq:hsdiag}))   $H^\text{NH}(t) = S(t) \left[D(t) - \frac{i}{4} \gamma_m \sigma_z\right] S^\dagger(t)$, where the anti-Hermitian part gives the upper (lower) level a negative (positive) imaginary part leading to an exponential decrease (increase) of the level in the time evolution, respectively.
This allows the NH time-evolution of $\rho(t)$ to relax the spin into its instantaneous ground state even without performing actual quantum jumps.
Thus, we expect qualitatively similar behavior of time evolution and energy transfer as compared with the full Liouvillian dynamics. This intuition is confirmed by the results of our numerical analysis shown in Fig.~\ref{fig:currents_eigenbasis_both}.
Here, the energy current into the bath can be shown to take the form
\begin{align}
   \dv{t}W_B^\text{NH}
   = \left(\dv{t} W_B^\text{L}\right) F(\rho, \rho^{\Downarrow})
\end{align}
which, as in the full Liouvillian case, vanishes in the limit of $\rho(t) = \ket{\Downarrow}\bra{\Downarrow}$.
This observation is in agreement with our numerical finding that the Lindblad and non-Hermitian dynamics exhibit full qualitative and even quite accurate quantitative agreement.
Thus, the effective NH Hamiltonian dynamics neglecting the jump terms provides a reasonable approximation for spin relaxation in the instantaneous eigenbasis.

\section{Spontaneous decay in a fixed basis}
\label{sec:fixed_basis}
We now discuss a dissipation channel in the form of spontaneous decay in a fixed basis which is modeled by the jump operator $L_i = \sigma_-$.
This scenario may correspond to a two-level system which microscopically has a large energy splitting $\omega_0$ and is subject to an inherent relaxation mechanism in this static energy eigenbasis. The driven coherent Hamiltonian $H_S$ is then understood in a frame rotating with frequency $\omega_0$, and thus defines a much lower energy scale in the spirit of a Jaynes-Cummings model within rotating wave approximation.
Again, we find a watchdog effect in the limit of strong coupling, this time pinning the spin to the time-independent state $\ket{\downarrow}$.
In addition, here the effective NH Hamiltonian approximation exhibits exceptional points which in themselves do not seem to have a striking qualitative effect on the dynamics of the system. However, we observe that the concomitant complex structure of the effective energy landscape, in particular line-degeneracies of the imaginary part known as imaginary Fermi-arcs are affecting the onset of non-adiabatic processes \cite{Bergholtz2021}.

\begin{figure}[htbp]
\includegraphics{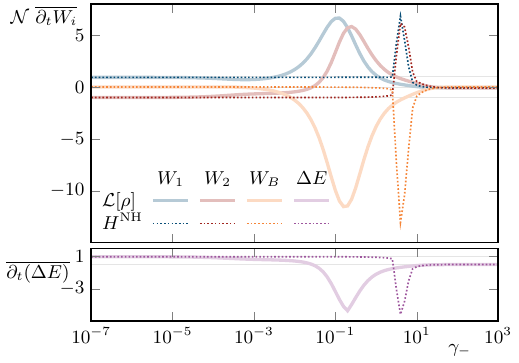}
\caption{
   Energy currents as a function of dissipation rate $\gamma_-$ for spontaneous decay in a fixed $\sigma_z$-basis.
   Top: Temporal average of the energy and bath currents $\mathcal{N}\pdv{W_i}{t}$ with $\mathcal{N} = \frac{2\pi}{\omega_1 \omega_2}$.
   Bottom: Temporal average of the effective energy transfer $\pdv{(\Delta E)}{t} = \frac{\mathcal{N}}{2} |\pdv{W_1}{t} - \pdv{W_2}{t}| - |\pdv{W_B}{t}|$.
   Note the significant differences between full Liouvillian (solid lines) and effective non-Hermitian dynamics (dotted lines) in all shown plots.
}
\label{fig:currents_fixed_both}
\end{figure}

\begin{figure}[htbp]
\includegraphics{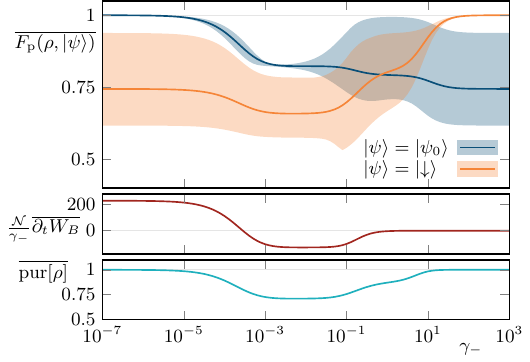}
\caption{
   Projected fidelities, $\gamma_-$-normalized bath current and temporal averaged purity as functions of the dissipation rate $\gamma_-$ for spontaneous decay in a fixed $\sigma_z$-basis in a fully Liouvillian description.
   Top: Temporal average of the projected fidelities (and 0.25- and 0.75-quantiles) of the state $\rho(t)$ with unperturbed state $\ket{\psi_0}$ and with instantaneous eigenstate of the Hamiltonian $\ket{\Downarrow}$.
   Center: Temporal average of the $\gamma_-$ normalized bath current $\frac{\mathcal{N}}{\gamma_-} \pdv{W_B}{t}$.
   Bottom: Temporal average of the purity $\pur[\rho] = \tr[\rho^2]$ of the system state $\rho(t)$.
}
\label{fig:results_Lindblad_fixed}
\end{figure}

\begin{figure}[htbp]
\includegraphics{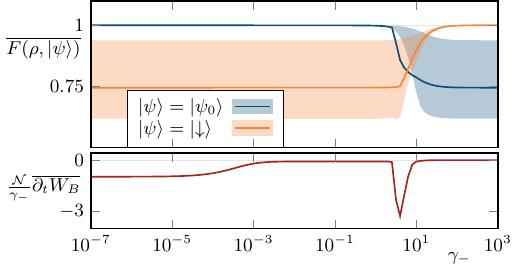}
\caption{
   Fidelities and $\gamma_-$-normalized bath current as functions of the dissipation rate $\gamma_-$ for spontaneous decay in a fixed $\sigma_z$-basis for the effective non-Hermitian description.
   Top: Temporal average of the fidelities (and 0.25- and 0.75-quantiles) of the state $\rho(t)$ with unperturbed state $\ket{\psi_0}$ and with instantaneous eigenstate of the Hamiltonian $\ket{\Downarrow}$.
   Bottom: Temporal average of the $\gamma_-$ normalized bath current $\frac{\mathcal{N}}{\gamma_-} \pdv{W_B}{t}$.
}
\label{fig:results_NH_fixed}
\end{figure}

Our numerical results are presented in Figs.~\ref{fig:currents_fixed_both}, \ref{fig:results_Lindblad_fixed} and \ref{fig:results_NH_fixed}.
The presented plots are similar in scope to the other illustrations regarding the shown quantities and parameter regimes.
In particular, in Fig.~\ref{fig:currents_fixed_both} we see a transition from the quantized topological frequency converter to a region where the energy is fully consumed by the bath (around the peaks of $W_i(t)$) to a vanishing of all the currents.
The Figs.~\ref{fig:results_Lindblad_fixed} and \ref{fig:results_NH_fixed} show that this transition is accompanied by a transition from $\rho(t) = \rho_0(t)$ to an intermediate state to $\rho(t) = \ket{\downarrow}\bra{\downarrow}$ where $\ket{\downarrow}$ is the time-independent lower eigenstate of $\sigma_z$ ($\sigma_z \ket{\downarrow} = -\ket{\downarrow}$).
Clearly, $\ket{\downarrow}$ does not provide any power transfer.
In Fig.~\ref{fig:results_Lindblad_fixed}, we plot the mean purity of the system state $\rho(t)$ in order to show that the competition of adiabatic dynamics and decoherence leads to a non-trivial mixing of both target states, as expected.

An interesting finding for spontaneous decay in a fixed basis is the appearance of an anomalous peak of the slope of $\frac{\mathcal N}{\gamma_-} W_B^\text{NH}(t)$ in Fig.~\ref{fig:results_NH_fixed} (at around $\gamma_- \approx 5$) that only shows up within the effective NH approximation but not in the full Liouvillian description.
The reason for its appearance can be found in non-adiabatic processes which are connected to the aformentioned imaginary Fermi arcs.
Specifically, the state $\rho(t)$, which follows the pure state $\ket{\psi_0}$ closely for smaller coupling strengths $\gamma_-$, suddenly shows transitions between the instantaneous eigenstates $\ket{\Downarrow}$ and $\ket{\Uparrow}$ to the one closer (w.r.t.~$\expval{\sigma_z}$) to $\ket{\downarrow}$.
As soon as these sudden transitions between the two instantaneous eigenstates occur, the energy entering the bath is increased by orders of magnitude, the reason being that these jumps cover a large distance on the Bloch sphere in a short time and thus collect a large dynamical phase, which leads to a step-like increase of the bath current.

Moreover, we find that these jumps can occur whenever the phase of the complex energy difference $E_+ - E_- = 2 \sqrt{\Re[\symbf\beta]^2 - \Im[\symbf \beta]^2 + 2i\Re[\symbf \beta]\cdot\Im[\symbf \beta]}$ of the NH Hamiltonian $H^\text{NH}(t) = \alpha(t)\mathds{1} + \symbf\beta(t) \cdot \symbf\sigma$ ($\symbf\beta \in\mathbb{C}^3$) changes its sign.
If we identify $\omega_i t + \phi_i$ as $k_i$, then $\vec k(t)$ is a ray in $k_x, k_y$ space.
The above phase changes its sign on a closed one-dimensional curve $\Re[\symbf\beta(k_x, k_y)]\cdot\Im[\symbf\beta(k_x, k_y)] = 0$ corresponding to the imaginary Fermi arc.
Whenever $\vec k(t)$ crosses such a one-dimensional curve and spends a sufficient time (compared to the coupling $\gamma_-$) within the region with the opposite-signed phase, a non-adiabatic process may appear.
This is because the time-evolution generated by the NH Hamiltonian exponentially favors the preferred eigenstate, i.e.~the eigenstate with positive imaginary part.

\begin{figure*}[htbp]
\includegraphics{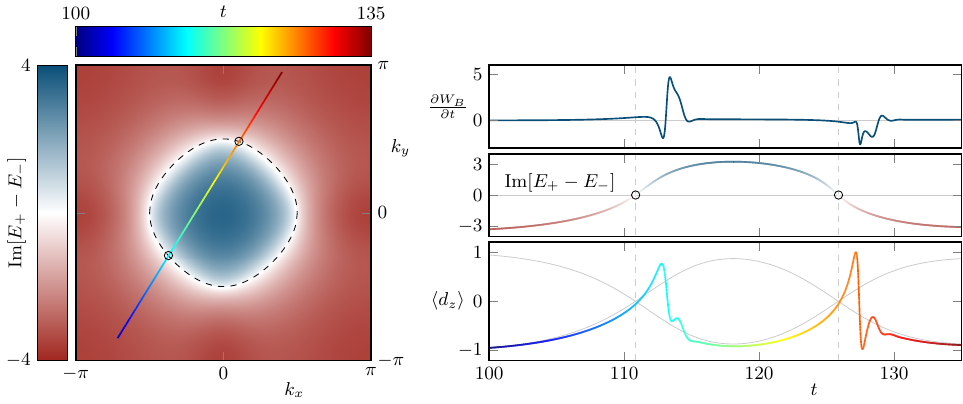}
\caption{
  Correspondence of non-adiabatic jumps between the instantaneous eigenstates of the Hamiltonian with the imaginary Fermi arcs for spontaneous decay in a fixed $\sigma_z$-basis for the effective non-Hermitian description.
  Left panel: imaginary part of the complex energy difference $\Im[E_+ - E_-]$ as function of $k_x, k_y$ (red-white-blue colors) and time-dependent phases $\vec\varphi_t$ (rainbow colors) in $(k_x, k_y)$-plane crossing the imaginary Fermi arc (dashed black line) defined by the curve $\Re[\symbf\beta(k_x, k_y)] \cdot \Im[\symbf \beta(k_x, k_y)] = 0$.
  Right panel (from top to bottom): bath current $\pdv{W_B}{t}$, imaginary energy difference $\Im[E_+ - E_-]$ and $z$-component of the Bloch vector $\expval{\sigma_z}$ of the state $\ket{\psi}$ as functions of time $t$.
  The circles in both panels mark the crossing of the imaginary Fermi arc.
  The parameters are chosen as in Sec.~\ref{ssec:numerical_approach} and $\gamma = 7$.
}
\label{fig:imaginary_energy_gap}
\end{figure*}

In Fig.~\ref{fig:imaginary_energy_gap} we show part of a state trajectory as it develops over time.
During this time evolution, the trajectory passes the aforementioned condition $\Re[\symbf\beta(k_x, k_y)]\cdot\Im[\symbf\beta(k_x, k_y)] = 0$ twice leading to two non-adiabatic jumps of the state.
The left panel of Fig.~\ref{fig:imaginary_energy_gap} shows the time-dependent phases $\vec\varphi_t$ as they evolve in the $(k_x, k_y)$-plane and the right panel shows the $z$-component of the Bloch vector $\expval{\sigma_z}$, the imaginary energy difference $\Im[E_+ - E_-]$ and the bath current $\pdv{W_B}{t}$ of the state $\ket{\psi}$.
One can clearly see the correspondence of the imaginary energy crossing and the non-adiabatic jumps of the state.
Note that in this case, the real part of the energy spectrum is gapped for all times, i.e.~$\Re[E_+ - E_-] > 1.9$, proving that this trajectory does not pass through an exceptional point but only crosses the imaginary Fermi arc.

The drastic effects of these non-adiabatic processes are a phenomenon exclusively found for individual (conditional on no jump) trajectories as described NH Hamiltonian dynamics, since the statistical ensemble $\rho(t)$ studied in the full Liouvillian description may reduce its purity so as to reduce the visibility of such processes. We note that the occurrence of imaginary Fermi arcs is a topological necessity in the presence of exceptional points, which thus provide a sufficient condition for the observed non-adiabatic jumps. However, exceptional points are not a necessary condition for this phenomenon as lines of imaginary degeneracies also appear in parameter regimes without exceptional points \cite{Shen2018,Zhou2018,Bergholtz2021}.

\section{Concluding discussion}
In summary, we have shown how dissipation in the form of spontaneous decay and dephasing affects the topological frequency converter and specifically its quantized power transfer.
We found that continuous deviations from the topologically quantized value may occur with the onset of dissipation. These deviations become noticeable at least when the dissipation strength is comparable to the coherent energy scale (level splitting). For even stronger dissipation, the power transfer completely vanishes due to quantum watchdog effects, even if the watchdog pins the state to perfectly adiabatically trace out a topological state manifold.

Interestingly, pure dephasing immediately leads to an exponential decay of the power transfer with time and thus, a vanishing of the long-time power transfer.
We have shown that this is due to the progressive loss of purity of the quantum state following from the combination of temporal driving with the recurring destruction of the off-diagonal elements.
We emphasize that this behavior may be specific for quasiperiodically driven system and has not been found in a corresponding system with two spatial dimensions. In particular, in Ref.~\cite{Hu2016} the half-BHZ model has also been investigated with pure dephasing in the instantaneous eigenbasis.
In contrast to the vanishing of the power transfer observed in our present work, there a stabilization of the quantized Hall response from pure dephasing has been found.
We attribute this difference to the temporal character of the (formal) Floquet lattice dimensions in our present system. The influence of dephasing in the intermediate case of a one-dimensional Floquet system with a single driving frequency remains an interesting direction of future work.

Finally, we elaborate on the relation of our results to the findings of Ref.~\cite{Nathan2019}, where a spin in a cavity coupled to a circularly polarized cavity mode and a circularly polarized driving mode has been discussed as an extended topological frequency converter model taking into account some sources of decoherence.
There, the topological frequency transfer has been expressed as an average increase in the number of cavity photons $\langle n \rangle$ connected to the Chern number $\mathcal C$.
The influence of cavity dissipation and spin relaxation is then modeled using the so-called Universal Lindblad equation \cite{Nathan2020a}.
In Ref.~\cite{Nathan2019}, dissipation can support a steady state with a topologically quantized photon emission rate by stabilizing fluctuations in the cavity photon number.
By contrast, our model can be understood as a classical limit of the cavity mode (keeping only the two-level system quantum), i.e.~the cavity state is replaced by a coherent state $\ket{\alpha}$ with $\hat a \ket{\alpha} = \alpha \ket{\alpha}$ and $\abs{\alpha}^2 \gg 1$. In this limit, fluctuations in the cavity photon number have no natural counterpart.

\section*{Acknowledgments}
We would like to thank Björn Trauzettel for discussions.
We acknowledge financial support from the German Research Foundation (DFG) through the Collaborative Research Centre SFB 1143 (Project-ID 247310070), the Cluster of Excellence ct.qmat (Project-ID 390858490), and the DFG Project No.~459864239.
Our numerical calculations were performed on resources at the TU Dresden Center for Information Services and High Performance Computing (ZIH).

\appendix
\section{Derivation of purity dynamics with Dephasing}
\label{app:purity_dephasing}
We derive Eq.~(\ref{eq:ddt_pur}) in the index notation
\begin{align}
   \dv{t}\pur[\rho]
   &= \dv{t} \tr[\rho^2] \\
   &= 2 \tr[\rho \dv{t} \rho] \\
   &= 2 \tr[\rho \left(-i[H, \rho] + \gamma_z \tilde\sigma_z \rho \tilde\sigma_z - \frac{\gamma_z}{2} \{\rho, \tilde\sigma_z^2\}\right)] \\
   &= 2\gamma_z \left(\tr[\rho \tilde\sigma_z \rho \tilde\sigma_z] - \tr[\rho^2]\right)
\end{align}
Using $\rho = \frac{1}{2}(\mathds{1} + \symbf{d}(t) \cdot \symbf\sigma)$ and $\tilde\sigma_z^2 = \mathds{1}$ one can see that the terms with exactly one $\mathds{1}$ vanish because a traceless Pauli matrix remains and the terms with exactly two $\mathds{1}$ give $\tr[\mathds{1}] = 2$ which also vanishes due to the opposite signs.
Thus the term above simplifies to (extracting the factor $\frac{1}{4}$)
\begin{align}
   \frac{2}{\gamma_z}\dv{t}\pur[\rho]
   &= \tr[(\symbf{d}(t) \cdot \symbf\sigma) \tilde\sigma_z (\symbf{d}(t) \cdot \symbf\sigma) \tilde\sigma_z] - \tr[(\symbf{d}(t) \cdot \symbf\sigma)^2] \\
   &= \tr[(\symbf{d}(t) \cdot \symbf\sigma) \tilde\sigma_z (\symbf{d}(t) \cdot \symbf\sigma) \tilde\sigma_z] - 2 \symbf d(t) \cdot \symbf d(t)
\end{align}
Inserting $\tilde\sigma_z = S^\dagger(t) \sigma_z S(t) = \symbf{k}(t) \cdot \symbf\sigma$ with $\abs{\symbf{k}(t)} = 1$ gives products of four Pauli matrices.
These can readily be evaluated
\begin{align}
   \sigma_a \sigma_b \sigma_c \sigma_d
   &= \left(\delta_{ab}\mathds{1} + i\varepsilon_{abe} \sigma_e \right) \left(\delta_{cd}\mathds{1} + i\varepsilon_{cdf} \sigma_f\right) \\
   &= \delta_{ab}\delta_{cd} \mathds{1} + i\delta_{ab} \varepsilon_{cdf} \sigma_f + i\delta_{cd} \varepsilon_{abe} \sigma_e \nonumber \\
   &\quad- \varepsilon_{abe} \varepsilon_{cdf} \left(\delta_{ef}\mathds{1} + i\varepsilon_{efg} \sigma_g \right) \\
   &= \left[\delta_{ab} \delta_{cd} - \varepsilon_{abe} \varepsilon_{cde}\right] \mathds{1} + \mathscr{O}(\sigma_i) \\
   &= \left[\delta_{ab}\delta_{cd} - \delta_{ac}\delta_{bd} + \delta_{ad}\delta_{bc}\right] \mathds{1} + \mathscr{O}(\sigma_i)
\end{align}
The terms which are proportional to a Pauli matrix $\mathscr{O}(\sigma_i)$ vanish because the Pauli matrix is traceless.
Thus, the final result is given by (the $\delta_{ij}$ terms become scalar products between the corresponding vectors)
\begin{align}
   \frac{2}{\gamma_z}\dv{t}\pur[\rho]
   &= 4\left(\symbf{d}(t) \cdot \symbf{k}(t)\right)^2 - 4\symbf d(t) \cdot \symbf d(t) \\
   &= 4\abs{\symbf{d}(t)}^2 \left(\cos(\angle{\symbf{d}(t), \symbf{k}(t)}) - 1\right)
\end{align}

Since $\tr[\rho^2] = \frac{1}{2}\left(1 + \abs{\symbf{d}(t))}^2\right)$ and $\cos(\angle{\symbf{d}(t), \symbf{k}(t)}) - 1 \leq 0$ we can see that indeed, $\abs{\symbf{d}(t)}^2$ is monotonically decreasing.

\section{Derivation of purity dynamics with spontaneous decay}
\label{app:purity_decay}
\begin{figure}[htbp]
\includegraphics[width=8.5cm]{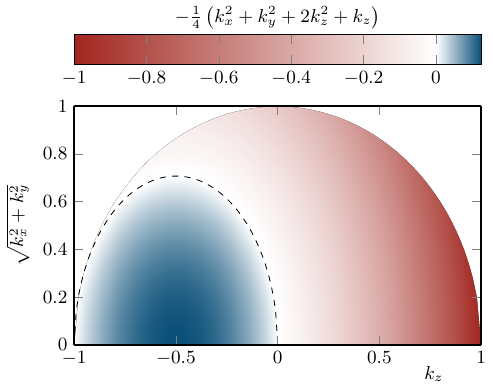}
\caption{
   The plot shows the the right hand side of Eq.~(\ref{eq:purity_decay}) for $\gamma=1$ and a countour line where it vanishes.
}
\label{fig:slope_purity_decay}
\end{figure}

Assume a Lindblad master equation with Lindblad operator $L_i(t) = S(t) \sigma_- S^\dagger(t)$ where $S(t)$ is a unitary transformation.
A similar calculation as performed in Appendix \ref{app:purity_dephasing} gives the purity (expressed by the length of the Bloch vector) as
\begin{align}
   \dv{t}\left(k_x^2 + k_y^2 + k_z^2\right) = -\frac{\gamma}{4}\left(k_x^2 + k_y^2 + 2k_z^2 + 2 k_z\right)
   \label{eq:purity_decay}
\end{align}
with $\symbf k \cdot \symbf\sigma = S^\dagger(t) (\symbf d \cdot \symbf\sigma) S(t)$.
The right side of the equation may take positive or negative values which would increase or decrease the purity of the state, respectively (see Fig.~\ref{fig:slope_purity_decay}).
Depending on the chosen basis this may lead to the purity fluctuating or converging to 0 or 1.
Indeed, the purity fluctuates for a fixed basis ($S(t) = \mathds{1}$) and converges to 1 in the instantaneous eigenbasis of the Hamiltonian ($H(t) = S(t) D(t) S^\dagger(t)$ with diagonal matrix $D(t)$).
The fluctuations arise because the \emph{target state} to which the system tries to converge is constantly changing.
In contrast, in the instantaneous eigenbasis, the target state is always the ground state of the Hamiltonian ($\symbf k = \left(0, 0, -1\right)$) and thus, the purity increases until this ground state is reached.

\section{State visualization for different coupling strengths in instantaneous eigenbasis}
\begin{figure*}[t]
    \includegraphics{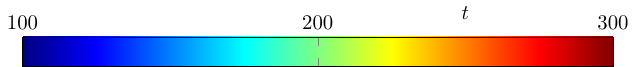} \\
    \vspace{0.3cm}
    \subfloat[$\gamma_- = 10^{-6}$]{%
    \includegraphics[width=0.32\textwidth]{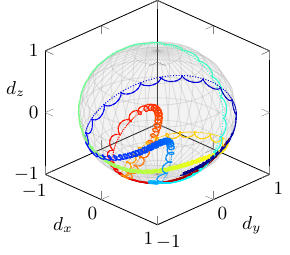}%
    }
    \subfloat[$\gamma_- = 0.1$]{%
    \includegraphics[width=0.32\textwidth]{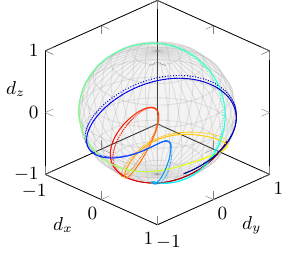}%
    }
    \subfloat[$\gamma_- = 100$]{%
    \includegraphics[width=0.32\textwidth]{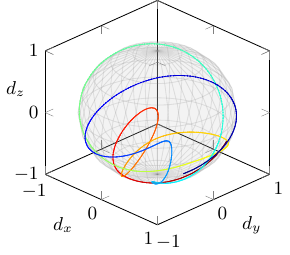}%
    }
    \caption{State of the system (solid line) and instantaneous eigenstate of the Hamiltonian (dotted line) as they evolve in time $t$ on the Bloch sphere for spontaneous decay in the instantaneous eigenbasis for different values of the dissipation strength $\gamma_-$.
    }
    \label{fig:states_on_blochsphere}
\end{figure*}

Fig.~\ref{fig:states_on_blochsphere} shows the evolution of the state $\rho$ as it evolves on the Bloch sphere for different values of $\gamma_-$ in case of spontaneous decay in the instantaneous eigenbasis.
One can see that, first ($\gamma_- = 10^{-6}$), the state oscillates around the instantaneous eigenstate of the Hamiltonian.
These oscillations are mainly responsible for the energy transfer of the frequency converter.
At intermediate interaction with the environment ($\gamma = 0.1$) these oscillations are damped and the state evolves in the close vicinity of the instantaneous eigenstate but still with a finite offset.
At very strong coupling ($\gamma = 100$), the state and the instantaneous eigenstate coincide.
Note that the values of $\gamma$ are chosen such that they fall into the plateaus visible in Fig.~\ref{fig:results_eigenbasis_Lindblad}.

For dephasing, the results are similar up to the additional effect of the shrinking of the Bloch sphere.

\section{Results for mixing of dephasing and spontaneous decay in instantaneous eigenbasis}
\label{app:results_gamma_md_d}
\begin{figure}[htbp]
\includegraphics[width=8.5cm]{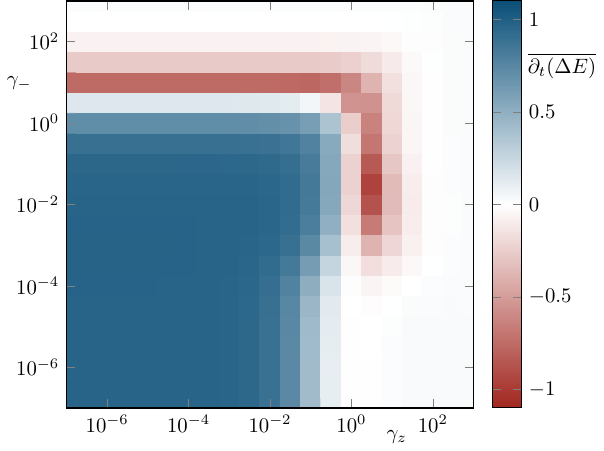}
\caption{
   Temporal average of the effective power transfer $\pdv{t}\Delta E$ as a function of the dissipation rates $\gamma_-$ (spontaneous decay in the instantaneous eigenbasis) and $\gamma_z$ (dephasing).
   The system is modelled within the full Lindbladian description and the parameters chosen are given in Sec.~\ref{ssec:numerical_approach}.
}
\label{fig:power_transfer_gamma_md_z}
\end{figure}

Fig.~\ref{fig:power_transfer_gamma_md_z} shows the temporal mean of the effective power transfer $\pdv{t}\Delta E$ (see Eq.~(\ref{eq:effective_energy_transfer})) as a function of a combination of the dissipation rates $\gamma_-$ (spontaneous decay in the instantaneous eigenbasis) and $\gamma_z$ (dephasing).
One can see that the spontaneous decay which favors the pure state $\ket{\Downarrow}$ restores the long-time energy transfer if the dephasing is small enough.
This indicates that the progressive loss of purity is reversed by spontaneous decay. \\

\section{Results for spontaneous decay in instantaneous eigenbasis in different topological regimes}
\begin{figure}[htbp]
\includegraphics[width=8.5cm]{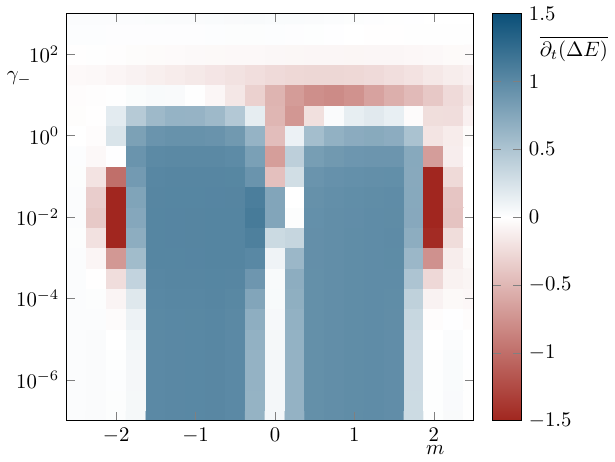}
\caption{
   Temporal average of the effective power transfer $\pdv{t}\Delta E$ as a function of the dissipation rate $\gamma_-$ for spontaneous decay in the instantaneous eigenbasis and the Zeeman parameter $m$ selecting the topology of the system.
   The system is modelled within the full Lindbladian description and the remaining parameters chosen are given in Sec.~\ref{ssec:numerical_approach}.
}
\label{fig:results_dissipation_topology}
\end{figure}

Fig.~\ref{fig:results_dissipation_topology} shows the temporal mean of the effective power transfer $\pdv{t}\Delta E$ (see Eq.~(\ref{eq:effective_energy_transfer})) as a function of the dissipation rate $\gamma_-$ (spontaneous decay in the instantaneous eigenbasis) and the Zeeman parameter $m$ (see Eq.~\ref{eq:tfc}).
For $0 < \abs{m} < 2$ the system is in a topological non-trivial phase with an additional phase transition at $m = 0$.
In Fig.~\ref{fig:results_dissipation_topology} one can clearly see the phase transitions and the topological quantisation of $\pdv{t}\Delta E$ within the topological phases.
Similar to Fig.~\ref{fig:currents_eigenbasis_both} one can see the transition of the power transfer from nearly quantized to zero due to the watchdog effect.

\label{app:results_decay_topology}

%

\end{document}